# Rheological properties
# of rotator and crystalline phases of alkanes

Diana Cholakova, Krastina Tsvetkova
Slavka Tcholakova, Nikolai Denkov*

*Department of Chemical and Pharmaceutical Engineering*
*Faculty of Chemistry and Pharmacy, Sofia University,*
*1 James Bourchier Avenue, 1164 Sofia, Bulgaria*

\*Corresponding author:
Prof. Nikolai Denkov
Department of Chemical and Pharmaceutical Engineering
Sofia University
1 James Bourchier Ave.,
Sofia 1164
Bulgaria
E-mail: nd@lcpe.uni-sofia.bg
Tel: +359 2 8161639
Fax: +359 2 9625643

**Abstract:**

Linear long-chain organic molecules are known to form lamellar intermediate phases (called also rotator phases) between their fully ordered crystalline phases and their isotropic liquid phases. The temperature range of occurrence and the properties of these intermediate rotator phases are crucially important for the texture of many food and cosmetic products, for the flow of various phase-change materials used in energy storage and transport, and for several processes in living nature, such as the formation of superhydrophobic and self-healing cuticle layers in plants and insects. Nevertheless, data for the rheological properties of such peculiar materials below their melting point, $T_m$, are almost missing, due to the specific difficulties in the respective rheological measurements. In the current study we describe a methodology for measuring and comparing the shear rheological properties of rotator (R) and crystalline (C) phases formed in bulk hydrocarbons at temperatures below $T_m$. We apply this approach to characterize the rheological properties of R and C phases formed upon cooling of alkanes with chain length varied between 17 and 30 carbon atoms. For comparison, we study also several alkane mixtures and one alkene with double bond at the end of its chain. The obtained results show that the storage and loss moduli of the rotator phases are *ca.* 10-times lower than those of the respective crystalline phases. We found also that the rheological properties of the crystal phases depend mainly on the subcooling temperature below the crystallization temperature, $T_C$, while the R phases become softer with the increase of the molecular length. We explain these results by assuming that the rheological properties of the crystal phases are determined mainly by the sliding of the ordered crystal domains with respect to each other, while in the rotator phases we have multiple defects in the molecular packing which increase with the alkane length. The proposed methodology and the obtained results serve as a solid basis for further rheological studies of this important class of technological systems.

# 1. Introduction

Normal alkanes with linear saturated chain ($C_n$) are the simplest organic molecules. Nevertheless, rich phase behavior has been observed upon cooling and heating of alkanes with intermediate chain length falling in the range between *ca.* 16 and 50 C-atoms and also for shorter odd-numbered alkanes [1-9]. Between their fully disordered liquid phases (L) and completely frozen crystalline phases (C), these alkanes form intermediate phases, called crystalline rotator phases (R). Molecules arranged in a crystalline rotator phase, called hereafter for simplicity just a "rotator phase", have long-range positional order in a lamellar structure, while their rotational freedom is not completely suppressed, *i.e.* the molecules in the R phase may rotate or oscillate with relatively large amplitude around their long axis. The R phases were found to be similar in some aspects to the smectic liquid crystalline (LC) phases which also have a lamellar structure, but the molecules in the LC phases have fluid-like behavior within the layers [5-7,10].

With respect to their thermodynamic stability, the rotator phases are divided into three groups – stable, metastable and transient [9,11]. The thermodynamically stable rotator phases form in alkanes with odd number of C-atoms and for even numbered alkanes when $n \geq 22$. In this case, the polymorphic behavior is of enantiotropic type, i.e. the R phases are with the lowest Gibbs energy in a given temperature interval and they form upon both cooling and heating. The metastable rotator phases are observed in the bulk $C_{18}$ and $C_{20}$ alkanes. These phases exist in a relatively narrow temperature interval and their free energy is higher than the free energy of the crystalline phase (monotropic polymorphism) – therefore, the metastable phases are observed upon cooling only, while they are not observed upon heating and melting of a crystal [11]. A very short-living, transient rotator phase is observed in the bulk $C_{16}$ alkane (hexadecane) upon cooling from melt – this phase was found to exist for a few seconds to several minutes only [11].

Five different rotator phases, denoted as $R_X$, where the subscript X is a Roman number between I and V, were identified in bulk alkanes with different chain lengths [6,7,11], see **Figure 1**. For example, short-chain alkanes (up to $C_{21}$) form only $R_I$ phase between their C and L phase; two consecutive phases are observed for $C_{22}$ (C → $R_I$ → $R_{II}$ → L) and $C_{28}$ to $C_{30}$ (C → $R_{III}$ → $R_{IV}$ → L) and the richest phase behavior with three R phases is detected for $C_{23}$ to $C_{26}$ alkanes. For $C_{23}$ to $C_{25}$ the phases are: C → $R_V$ → $R_I$ → $R_{II}$ → L, whereas they are C → $R_V$ → $R_{II}$ → $R_{IV}$ → L for $C_{26}$. Longer chain length alkanes ($n \geq 31$) form only $R_{III}$ phase. The distinct R phases differ from one another by the type of their unit cells, layer sequence and molecular tilt with respect to the layers [6,7], see **Figures 1b** and **1c**.

The rich phase behavior of alkanes includes also several different crystalline lattices which form after the R-to-C phase transition [4,12]. The even-numbered alkane with $n$ between 14 and 24, as well as the odd-numbered alkane with $n < 13$ crystallize in triclinic lattice; even-numbered alkanes with $26 \leq n \leq 38$ form monoclinic lattice, whereas the odd-numbered alkanes with $13 \leq n \leq 41$ and even-numbered alkanes with $40 \leq n \leq 66$ arrange in orthorhombic crystalline lattice [12].

Beside in bulk alkanes, rotator phases have been detected in surface layers of alkanes and in alkanes confined in micro- and nano-structures, such as pores, emulsion drops and polymeric capsules [9,13-21]. Furthermore, rotator phases have been observed also with long-chain alkenes, alcohols, asymmetric alkanes and their mixtures [9,22-29].

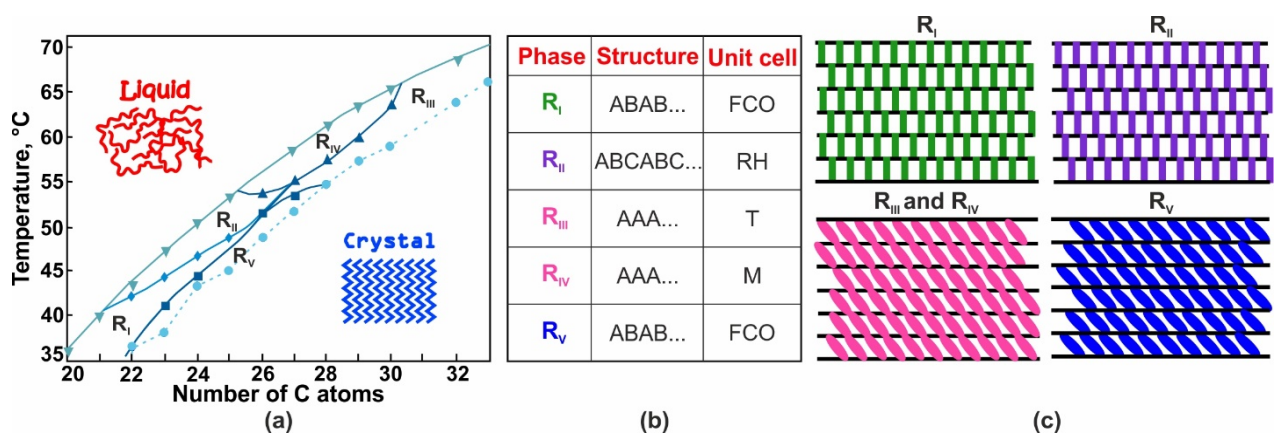

**Figure 1. Rotator phases.** (a) Phase diagram of the R phases observed upon cooling of alkanes with chain length between 20 and 33 C-atoms. (b) Main structural characteristics of the various rotator phases. The unit cells are: FCO – face centered orthorhombic; RH – rhombohedral; T – triclinic and M – monoclinic. (c) Schematic presentation of the molecular ordering within the layers in the various R phases. The difference between $R_{III}$ and $R_{IV}$, which have the same representation in this schematic, is the direction of tilt of the molecules within the layer: in $R_{III}$ the molecules are tilted toward their nearest neighbors, whereas in $R_{IV}$ the tilt direction is toward the next-nearest neighbors. Adapted from Refs. [6,9].

The rheological properties of rotator phases are important for many technological applications and for various processes in living nature. For example, the petroleum jelly and paraffin wax contain a mixture of alkane molecules in R phase – they are used in many cosmetic products, because of their unique ability to easily spread in continuous soft and self-healing layers, which is due to the specific rheological properties of the respective R phase [30-35]. One type of energy storage and transportation systems uses aqueous dispersions of alkane micro-particles as phase change materials (PCMs) due to their high latent heat and chemical stability [36-40]. In this case, the dispersions should be able to flow freely in the whole temperature interval needed for a particular application. The formation of R phases in the alkane particles, upon heating and cooling of the dispersions, turned out to be a very important factor in the selection of appropriate alkane-surfactant combinations. Alkanes and their mixture are also widely used as lubricants in industry which is also related to the peculiar rheological properties of the lamellar phases formed under the operational conditions [41,42].

In addition, several studies have shown that the phase behavior of various waxes (containing usually a mixture of long-chain alkanes, alcohols, fatty acids, glycerides, phospholipids and glycolipids) is very important for some processes in living organisms [9,43-47]. For example, the poor wetting of the surface of insects by water drops is determined by the peculiar structure and properties of a thin epicuticular layer of wax. This layer prevents also the trans-cuticular water loss, viz. the dehydration of the insect body [9,43,44]. Another interesting example for the importance of rotator phases in natural environment is the phase behavior of beeswax [45,46]. The crystal to rotator phase transition in beeswax takes place at *ca.* 36.3°C, which is exactly the temperature maintained in the honeycomb construction zone. It has been proposed that the natural softness of the rotator phases is used by the bees to construct the specific hexagonal shapes of their honeycombs [45]. Finally, we note that the myelin and chromatin contained in the nucleus of the cerebral cells in rats are known to contain alkane rotator phases as well [9,47]. For these rotator phases, the C-to-R transition takes place at about 30°C, whereas the R-to-L phase transition occurs at 44°C which coincide with the upper temperature limit for rats functioning as living organisms.

Despite the known multiple examples for the importance of the rheological properties of rotator phases, these properties remain poorly understood. Several studies [41,48-52] have explored the visco-elastic properties of various paraffin waxes which contain mixtures of alkane molecules, as explained below.

Wang et al. [48] measured the dynamic shear moduli of paraffin wax as a function of temperature. They reported values for the storage modulus ($G'$) of about 700 MPa at temperatures -40 to 0°C (ordered phase), whereas the modulus was more than three orders of magnitude lower, ≈ 0.25 MPa, in the intermediate mesophase region at temperatures between 40 and 50°C (where the wax molecules are ordered in rotator phase). The loss modulus ($G''$) had similar magnitude to $G'$ in the rotator phase region, whereas it was ≈ 30 MPa in the crystalline zone.

Petersson and co-authors [49] reported significantly lower shear moduli for microwax: for crystalline phase at 10°C they measured $G' ≈ 9.5$ MPa and $G'' ≈ 0.15$ MPa at low strain. Upon heating the elastic modulus decreased almost linearly its value from *ca.* 9 MPa at 25°C to 5 MPa at 50°C. A slight linear increase of $G''$ upon temperature increase was measured.

Nowak and Severtson [52] measured the dilatational moduli of commercial wax using dynamic mechanical spectroscopy (DMS) method. In this study, the dilatational moduli were $E' ≈ 30$ MPa for the crystalline phase and around 12 MPa for the rotator phase, whereas $E''$ was around 2-5 MPa for both C and R phases. The properties of pure heneicosane ($C_{21}$ alkane), tricosane ($C_{23}$ alkane) and their 79:21 mol/mol mixture were also measured in the same study. The storage modulus for pure $C_{21}$ and $C_{23}$ was ca. 55 MPa in C phase and decreased down to 45 MPa for $R_I$ phase of $C_{21}$ and 33 MPa for the same phase of $C_{23}$. The loss modulus was several times lower, *ca.* 10 MPa for $C_{21}$ and 20 MPa for $C_{23}$ in the C phase, and 11 and 17 MPa for the respective $R_I$ phase. For $R_{II}$ phase observed with $C_{23}$, $E' ≈ 17$ MPa and $E'' ≈ 9.5$ MPa were measured. The $R_V$ phase, which is the closest one to the C phase in $C_{23}$, was not detected in these studies, most probably because of the relatively high heating rate of the experiments, 5°C/min. Furthermore, most probably due to limited time for experiments at 5°C/min, the authors concluded that the "plastic crystalline states of *n*-alkanes are difficult to isolate" and that the "observed viscoelasticity surrounding the mesophases is in great part due to transitions between phases, not the rotator phases themselves" [52].

Several studies explored also the viscosities of the melted *n*-alkanes [53,54]. Other studies reported computational results about the rheological properties of alkanes calculated in molecular dynamics simulations, however, no comparison with relevant experimental data is currently available [55-57].

This short summary of the rheological results available for the rotator phases of alkanes shows a large gap in our knowledge and understanding of this important class of systems. The first major aim of the current study is to define a robust rheological procedure for detection of the phase transitions in alkanes upon cooling and heating around their melting point. The second major aim is to use this procedure for systematic characterization of the viscoelastic properties of the crystalline and rotator phases of alkanes with chain length varied between 17 and 30 C-atoms, of several alkane mixtures and one linear alkene. The obtained results allow us to define several general trends and to propose their plausible structural explanations.

## 2. Experimental materials and methods

### 2.1 Materials

We studied linear alkanes with chain length varied between 17 and 30 C-atoms, denoted in the text as $C_n$. All rotator phases studied are thermodynamically stable and form upon both cooling from melt and heating from crystal. The bulk melting temperatures ($T_m$) as reported in the literature, the chemical purity and their producers are described in **Supplementary Table S1**. We studied also the phase behavior of one alkene with single double bond at α-position, 1-eicosene ($C_{20:1}$, product of TCI Chemicals, purity > 97%). The alkane mixtures used in some series of experiments were prepared by mixing pre-melted alkanes and homogenization of the mixture in a liquid state. All alkane ratios presented in the manuscript are molar (mol/mol).

### 2.2 Methods

#### 2.2.1 Rheological measurements

The rheological properties were measured with rotational rheometers Discovery Hybrid Rheometer DHR-3 and Discovery Hybrid Rheometer HR-20 (TA Instruments, USA). We used parallel-plates geometry with upper plate of 40 mm in diameter. The gap between the two plates was set to 300 μm. Prior to the measurements we glued ultra-fine sand paper of grade P2000 on both the lower and the upper plates to prevent a possible wall slip between the sample and the plates during the experiments [58].

The rheological response of the samples was characterized in oscillatory deformations upon temperature ramps with cooling/heating rates varied between 0.1 and 3°C/min. In most of the experiments 0.5°C/min rate was used. In all experiments, the tested sample was placed in the rheometer at high temperature ($T > T_m$), ensuring that it was in a liquid state, and the temperature was equilibrated for 180 s. Afterwards, the shear experiments upon cooling were started. For the heating experiments, we first cooled the liquid sample to the desired low temperature and then the heating was started with the same temperature rate as the one used for cooling. The shear storage modulus ($G'$) and the shear loss modulus ($G''$) were measured as a function of temperature at fixed strain, γ = 0.05%, and frequency, 1 Hz. These parameters were chosen after performing a preliminary series of tests for determining the best experimental conditions, see **Supplementary Figures S1** and **S2** in **Supplementary** **materials**. Results from at least 3 independent experiments were averaged and used for preparation of the graphs shown in this paper. The error bars in the figures represent the standard deviations, calculated among all results obtained under equivalent experimental conditions. The notation of the rotator phases presented within the graphs is based on the literature data [6]. No direct structural characterization was performed in the current study.

### 2.2.2 Differential scanning calorimetry (DSC)

DSC experiments were performed to verify the temperature ranges of the phase transitions in our samples under equivalent cooling and heating rates. We used Discovery DSC 250 apparatus (TA Instruments, USA). A small amount of the studied sample (2-10 mg) was placed in aluminum hermetic pan, weighted and hermetically sealed with aluminum hermetic lid using Tzero sample press. The samples were then placed in the DSC oven and the following measurements protocol was applied: (1) The sample was heated from 30°C to temperature which is by 10°C higher than the bulk melting temperature of the studied sample; (2) The temperature was kept constant for 5 min to ensure the complete sample melting; (3) The sample was cooled to temperature which is at least 10°C lower than its freezing temperature; (4) The samples were again heated to $T = T_m + 10$°C. The DSC curves presented throughout the paper are the first cooling and the second heating curves (after the first cooling) as defined in the current procedure, *i.e.* obtained in steps (3) and (4). Fixed rate of 0.5°C/min was used for these two steps. The initial heating (step 1) was performed at 10°C/min rate.

The obtained thermograms allow us to determine the phase transitions temperatures and enthalpies. Their analysis was performed using the build-in functions of the Trios data analysis software (TA Instruments, USA).

### 2.2.3 Microscopy observations

The optical observations were performed with AxioImager.M2m microscope (Zeiss, Germany). Long-focus objectives ×20 and ×50 were used in transmitted and cross-polarized white light with included compensator plate, situated after the sample and before the analyzer at 45° angle with respect to both the polarizer and analyzer.

## 3. Results and discussion

In the current section we present the obtained results is the following order: First, we show illustrative results for heneicosane ($C_{21}$), tricosane ($C_{23}$) and hexacosane ($C_{26}$) obtained via the defined rheological procedure. These results show that we can detect the L→$R_X$, $R_X \rightarrow R_Y$ and/or $R_Y \rightarrow C$ phase transitions during the cooling/heating cycle for a single alkane. Next, we compare the results obtained with all studied alkanes and discuss how the specific phase (type of crystalline/crystalline rotator phase) affects the measured rheological properties. Finally, we show that the same approach can be used to study the rheological properties of rotator phases of other materials and compare the results obtained with pure alkanes to those obtained with alkane mixtures.

### 3.1 Rheological properties of heneicosane ($C_{21}$)

The heneicosane, $C_{21}$, exhibits relatively simple phase behavior. Upon cooling from melt, $R_I$ phase is observed only before crystallization in the thermodynamically stable orthorhombic crystalline phase. The DSC and rheological curves obtained upon cooling of this alkane are presented in **Figure 2a,b** and the curves obtained in heating experiments – in **Figure 2c,d**. As seen from the thermograms, two peaks are observed upon both cooling and heating showing the two phase transitions which occur in this sample, $L \rightarrow R$ and from $R \rightarrow C$ upon cooling and the reverse sequence upon heating. The enthalpies of these phase transition are $\Delta H_{LR} \approx 156.8$ J/g and $\Delta H_{RC} \approx 55.8$ J/g, respectively. They are in a good agreement with the data reported in the literature: $\Delta H_{LR} \approx 160.9$ J/g and $\Delta H_{RC} \approx 52.2$ J/g [4]. We note that the curve reversal observed in the $R \rightarrow C$ peak in the $C_{21}$ cooling thermogram is due to the instant sample overheating, due to the heat released upon the rotator-to-crystal exothermic phase transition. This transition causes instant temperature increase which is compensated by the instrument within a few seconds and the cooling continues. Similar DSC loops were observed also in the other studied alkanes, see for example **Supplementary Figure S4**.

The same phase transitions are observed also in the rheological experiment – see the stepwise increase of the measured moduli in the rheological curves, **Figure 2a,b**. The liquid alkane has low viscosity and flows freely; therefore, the respective storage and loss moduli, $G'_L$ and $G''_L \approx 0.1 \div 1$ Pa, are in the range of the rheometer detection limit. Significant increase by about 5 orders of magnitude is observed during the $L \rightarrow R$ phase transition upon cooling, $G'_{RI} \approx 0.11$ MPa and $G''_{RI} \approx 0.03$ MPa. Upon further cooling, the moduli increase steadily prior to the crystallization step, where a second stepwise increase of the moduli is observed. The magnitude of this second step is much smaller compared to the first one - the moduli increase by about one order of magnitude, thus achieving values of 2 MPa for $G'_C$ and $\approx 0.9$ MPa for $G''_C$. The relatively wide temperature intervals in which the phase transitions are observed in the rheological curves are most probably caused by two factors:

first, the sample needs some time to transfer fully from one phase to another and, second, the shearing force applied to the sample may induce a crystal-melt coexistence for a given period of time [59].

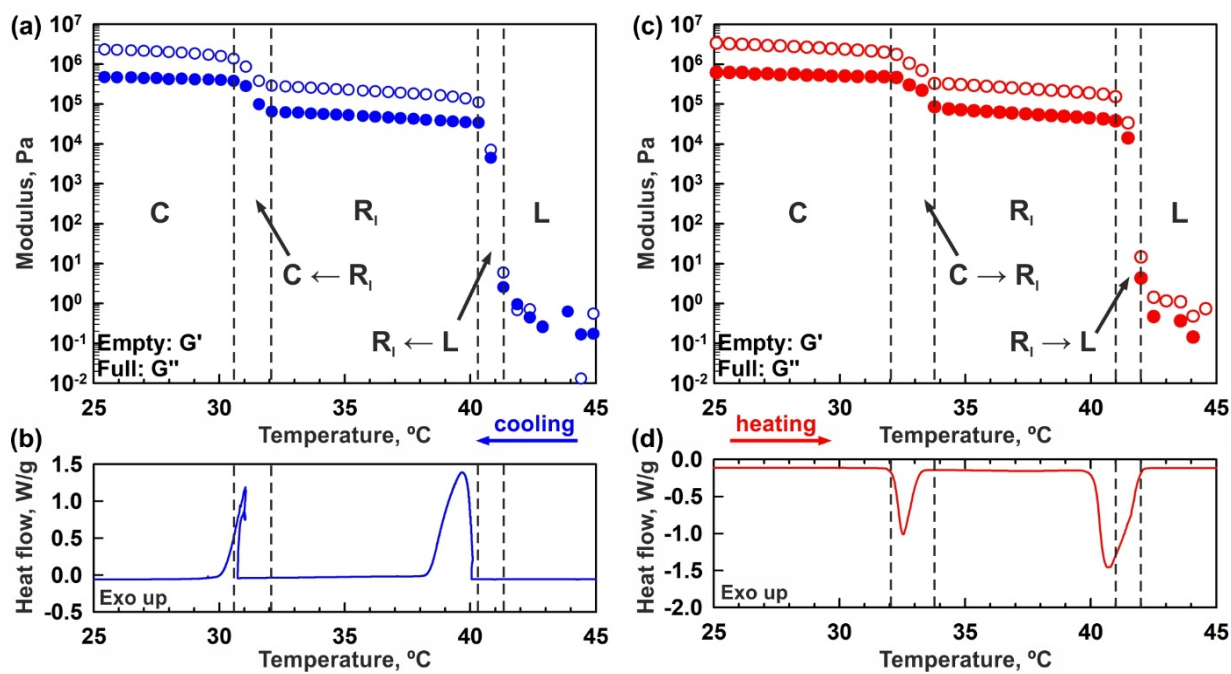

**Figure 2.** **Rheological (a,c) and DSC (b,d) results obtained with $C_{21}$ alkane** upon (a,b) cooling and (c,d) heating. Dashed lines denote the temperature intervals in which the L → $R_I$ and $R_I$ → C phase transitions are detected in the rheological experiments; see the main text for more explanations. All experiments are performed at 0.5°C/min rate. Empty symbols show the storage modulus ($G'$) and the full symbols – the loss modulus ($G''$).

The moduli measured during the cooling and heating experiments were in excellent agreement with each other. This comparison shows that the shearing with small amplitude, applied upon cooling, does not affect significantly the formed crystalline structure and the respective rheological properties of the alkanes.

Comparison of the phase transition temperatures shows, however, a slight difference between the cooling and heating experiments, see **Supplementary Table S2** for summary of these temperatures. In the heating experiment, excellent agreement is observed between the DSC measurements and the rheological measurements, **Figure 2c,d**. The C→R phase transition is observed between 32.3 and 33.8°C in the rheological measurements, whereas the respective DSC peak is detected between 31.7 and 33.5°C with maximum at 32.6°C. The R→L transition begins at *ca.* 41°C in the rheological measurements, slightly after the peak maximum observed in the calorimetry experiments (39.7-42.3°C temperature range with 40.7°C peak maximum).

As expected, the temperatures of these phase transitions are slightly lower in the respective cooling experiments because of the time needed for formation of rotator or crystalline nuclei which are able to induce the molecular ordering, due to the known subcooling typically observed in the crystallization experiments [60,61]. The L→R phase transition is observed between 41.3 and 40.3°C in the rheological experiments, whereas the DSC peak has maximum 39.7°C (T range 40-39.2°C). Similar deviation is observed also for the second phase transition, R→C: it occurs between 32.1 and 30.6°C in the rheology experiments and between 30.8 and 29.8°C in the DSC experiments. The reason for the lower temperatures measured upon cooling in the calorimetry experiments (compared to the rheological measurements) is the very different amount of sample used in the two methods: in the rheological experiments we use samples of ≈ 300 mg, whereas the sample in the DSC experiments is a few milligrams only. As the probability for nucleation depends on the volume of the sample, slightly larger subcooling is expected when smaller sample is tested [60,61].

The results from these experiments demonstrate for the first time that the rheological measurements can be used for detection of the main phase transitions (L→R and R→C) in alkane samples when appropriate experimental protocol is used. Furthermore, the defined procedure has excellent reproducibility, see the experimental data obtained in several independent experiments with $C_{19}$ and $C_{21}$, compared in **Supplementary Figure S3**. The method allows us to compare results from experiments with different cooling and heating rates. Cooling rates between 0.5°C/min and 5°C/min were tested for crystallization of the samples used in the heating rheological experiments. The measured moduli upon heating were similar for samples cooled with up to 2°C/min cooling rate, see **Supplementary Figure S2**. Only when relatively high cooling rates were applied, e.g. 5°C/min, the measured moduli decreased slightly as previously reported [62,63]. To have equivalent experimental conditions in all other heating experiments, we cooled the sample always at 0.5°C/min, unless otherwise specified explicitly.

Next, we show experimental results for alkanes exhibiting more complex phase behavior, for which several distinct rotator phases are observed upon cooling.

## 3.2 Rheological properties of tricosane ($C_{23}$) and hexacosane ($C_{26}$)

Increasing the alkane chain length by 2 C-atoms only (from $C_{21}$ to $C_{23}$) leads to significantly richer phase behavior. The molecules of $C_{23}$ and $C_{26}$ can arrange in three different rotator phases which are observed between their isotropic liquid and fully crystalline phases. The exact phase transition sequence observed upon cooling is: L → $R_{II}$ → $R_I$ → $R_V$ → C for $C_{23}$ and L → $R_{IV}$ → $R_{II}$ → $R_V$ → C for hexacosane [6].

To check whether the developed rheological procedure can be used for detection of the phase transitions among the various rotator phases, we performed experiments with these two alkanes. The respective results from the rheological measurements are presented in **Figure 3** and the related DSC curves are shown in **Supplementary Figure S4**.

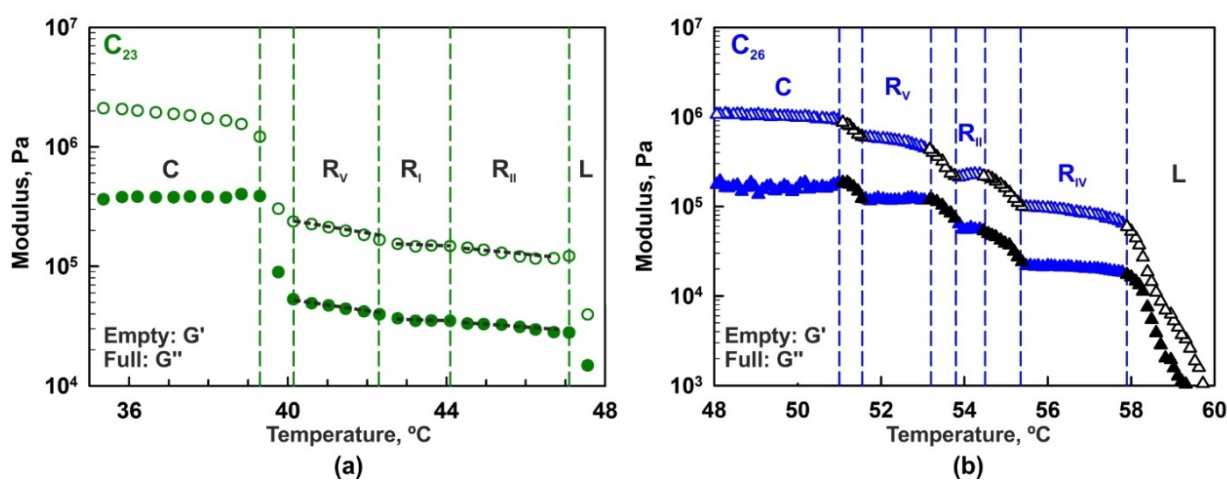

**Figure 3. Rheological curves obtained in cooling experiments with $C_{23}$ alkane (a) and $C_{26}$ alkane (b).** (a) When cooled $C_{23}$ molecules arrange in three different R phases prior to final crystallization. (b) Three different R phases are detected for $C_{26}$ sample as well. The black symbols represent the regions in which phase transitions between the distinct phases occur, whereas with blue symbols we show the moduli for a given phase. Note that the y-scale begins from $10^3$ or $10^4$ Pa for clarity – the points for the initial liquid phase are not shown. The experiments are performed at 0.5°C/min cooling rate. Empty symbols show the storage modulus ($G'$) and the full symbols – the loss modulus ($G''$). For more detailed explanations, see the main text.

The phase transitions with the highest enthalpy, L→R and R→L, occurred in a similar mode as those observed with $C_{21}$ sample. Two stepwise increase of the moduli to $10^4$-$10^5$ Pa upon the formation of the first R phase and to ≈ $10^6$ Pa for C phase were observed. Upon $R_X$ → $R_Y$ phase transition, the viscoelastic properties of the alkanes changed slightly but detectably. These transitions occurred with much smaller but detectable stepwise increase of the moduli or with small changes in the slope of the curves $G'(T)$ and $G''(T)$, see **Figure 3**. The first phase observed upon cooling of tricosane is $R_{II}$. The molecules are untilted with respect to the layers in this phase and the layers are packed in a hexagonal lattice with trilayer stacking sequence (also called "rhombohedral") [6]. Upon further cooling, a first order phase transition takes place and the molecules rearrange in $R_I$ phase, forming bilayer stacking sequence and rectangular distorted hexagonal lattice, referred also as "face centered orthorhombic" [6]. Afterwards, $R_V$ phase forms which has similar characteristics to $R_I$ but with tilted molecules. The $R_I$ to $R_V$ phase transition was initially characterized as being of the second order [6], but more recent studies with photopyroelectric calorimetry suggested that this is a first order phase transition [64]. The molecular structure with lowest energy for this alkane observed at low temperatures is orthorhombic.

The hexacosane is particularly interesting as this alkane is with the longest chain which exhibits $R_{II}$ phase (typical for shorter alkanes) being also the shortest alkane which forms $R_{IV}$ phase observed with $C_{26}$ to $C_{30}$ alkanes [6]. Furthermore, the X-ray experiments showed that $R_I$ phase which is typically observed as an intermediate between $R_{II}$ and $R_V$ is not observed for $C_{26}$ or it has "imperceptibly small temperature range" of existence [6].

Four distinctive steps showing the transitions proceeding in the hexacosane sample were observed in the $G'(T)$ and $G''(T)$ rheology curves. According to the literature, these transitions are: L → $R_{IV}$ → $R_{II}$ → $R_V$ → C. We note that we observed also a small slope change in the rheology curves at ca. 57°C and also a double initial peak at high temperature in the DSC measurement, see **Supplementary Figure S4c,d**. These results can be also interpreted as showing two consecutive phase transitions from L to $R_{IV}$ and then to $R_{II}$. The three steps observed at lower temperatures afterwards may be due to the formation of $R_I$ at $T \approx 54.5°C$, $R_V$ at $T \approx 53.2°C$ and C at $T \approx 51°C$ (temperatures determined from the rheological measurements upon cooling). To confirm or reject the possible formation of $R_I$ phase in $C_{26}$, however, we need further structural measurements, possibly coupled with shear oscillations (Rheo-SAXS). Such structural investigations go beyond the scope of the current study. For that reason, on our graphs we display the notation used for $C_{26}$ alkane in the literature [6,8,65].

The moduli of the rotator phases observed in $C_{23}$ and $C_{26}$ alkanes do not remain constant upon temperature decrease – their values increased steadily upon cooling, as in the case of heneicosane. This trend shows that the moduli of the R phase do not depend on the type of the molecular arrangement only but also on the specific temperature of measuring, compared to the temperature of phase formation, viz. on the subcooling of the phase. This trend is discussed further in **Section 3.3** below.

The presented experimental results show that the proposed experimental procedure can be used for successful detection of the phase transitions occurring in bulk alkane samples.

Until now, these transitions have been detected by X-ray [1-3,5-7,11,18,24], calorimetry techniques [3,8,16,18,24,64,65], infrared spectroscopy [2,3], Raman spectroscopy [3,66] and recently by dynamic light scattering [67].

### 3.3 Comparison between different alkanes

We performed similar rheological and DSC measurements for most of the alkanes with chain length between 17 and 30 C-atoms, see **Section 2.1** above. The obtained curves were qualitatively similar and, in all cases, we were able to detect the phase transitions among the different R phases. In the following two subsections we compare the obtained results from these experiments, aiming to define some general trends in the rheological properties of the alkane crystalline and rotator phases.

### 3.3.1 Crystal phases

The temperature at which the phase transition C→R (or R→C) occurs depends strongly on the alkane chain length. To compare the results for different alkanes, we used as a reference point the temperature at which the C→R transition for a given alkane occurs upon heating, $T_{CR}$. Thus we defined the temperature difference, $\Delta T_{CR} = (T - T_{CR})$, as the temperature at which the moduli are measured minus the temperature $T_{CR}$. Therefore, lower $\Delta T_{CR}$ values (higher in absolute value) correspond to larger subcooling and lower thermal energy of the molecules.

The comparison between the measured moduli for the crystalline phases of the various alkanes is presented in **Figure 4** as a function of $\Delta T_{CR}$, as determined in the heating experiments. Interestingly, using this simple scaling we obtained two master lines for the two moduli, as measured with all alkanes having chain-lengths between 19 and 28 C-atoms. Slight deviation to lower values was observed for $C_{22}$ only at $|\Delta T_{CR}| \leq 2°C$. The two moduli decreased linearly with the increase of temperature, while the results for the different alkanes fell around the following master lines within the frame of our experimental accuracy:

$$G'(\Delta T_{CR}) \approx - 0.219\Delta T_{CR} + 0.75 \text{ MPa} \qquad \text{(eq. 1)}$$

$$G''(\Delta T_{CR}) \approx - 0.037\Delta T_{CR} + 0.26 \text{ MPa} \qquad \text{(eq. 2)}$$

Subcooling, $\Delta T_{CR}$, of down to -8°C were tested. Similar, although slightly more scattered graphs were obtained using the experimental data from our cooling experiments, see **Supplementary Figure S5**. The master lines describing the moduli of the crystalline phases measured upon cooling are: $G'(\Delta T_{CR}) \approx -0.213\Delta T_{CR} + 0.75$ MPa and $G''(\Delta T_{CR}) \approx -0.055\Delta T_{CR} + 0.25$ MPa. For these sets of data, the average $G'/G''$ ratio was $\approx 3.7 \pm 0.8$.

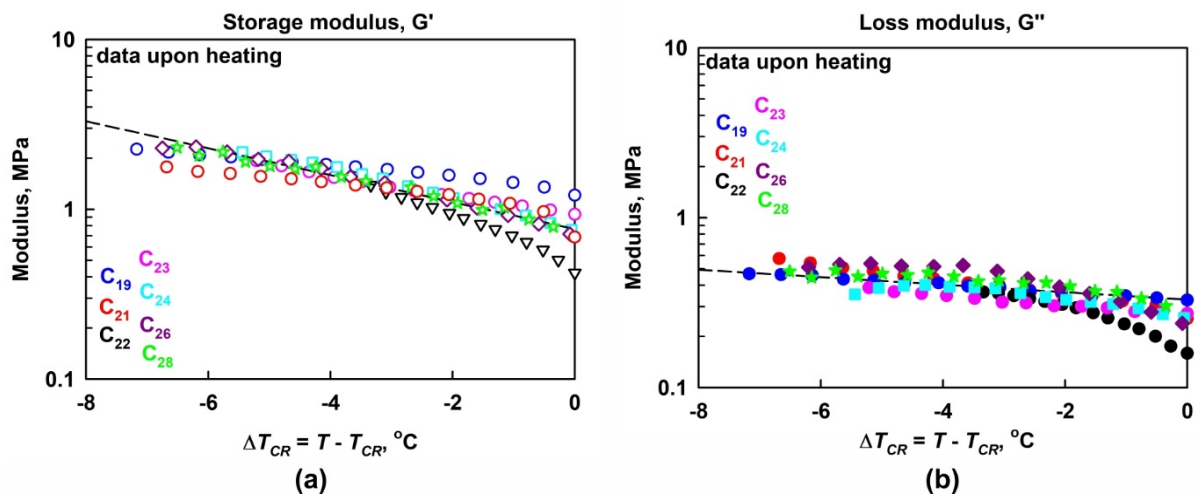

**Figure 4.** **Storage (a) and loss (b) moduli of crystalline alkane phases.** The moduli are plotted as a function of the subcooling with respect to the temperature at which the C→R phase transition occurs upon heating. Data for all alkanes with $n \geq 19$ lay on master lines, showing that the shear moduli do not depend significantly on the type of the crystal lattice formed. The same data points are presented with error bars in **Supplementary Figure S6**.

This result was quite unexpected, having in mind that different crystalline lattices are formed by these alkanes as revealed by X-scattering techniques [4,12]. As explained in the introduction, the alkanes form three crystalline lattices depending on their molecular lengths. Representatives of all three were tested: $C_{22}$ and $C_{24}$ crystallize in triclinic lattice, $C_{26}$ and $C_{28}$ in monoclinic, whereas the odd-numbered alkanes ($C_{19}$, $C_{21}$ and $C_{23}$) form orthorhombic cell [12].

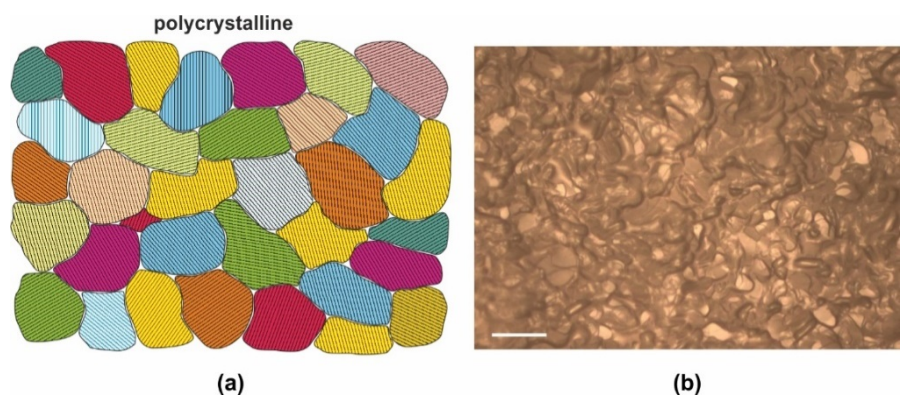

**Figure 5.** **(a)** Schematic presentation of the polycrystalline structure formed upon alkane crystallization. The distinct domains slide with respect to each other upon shear deformation. **(b)** Optical microscopy image illustrating the structure of bulk nonadecane at 27°C after shearing experiment. Scale bar = 50 μm.

The obtained results show unambiguously that the type of unit cell and the alkane chain-length do not affect strongly the shear rheological properties of the crystalline alkanes.

Therefore, there should be a common property of all alkanes, which determines the measured moduli. Upon crystallization, two possible scenarios could be hypothesized – either all molecules arrange without defects in a single monocrystalline structure or numerous small crystallites form and the sample has a polycrystalline structure. In the rheological measurements, the amount of sample is relatively large (≈ 300 mg). Therefore, one could expect that a polycrystalline structure would form upon crystallization. Indeed, our optical microscopy observations in polarized light confirmed that a polycrystalline structure is formed in these samples, see **Figure 5b**. The mechanically weakest regions in a polycrystalline structure, however, are usually not the lattices of the individual crystallites but the contact regions between the grains. Therefore, the most probable explanation of the obtained results is that the measured moduli show what is the relative friction between the separate crystalline domains and how easy is to slide one domain with respect to its neighbors, see **Figure 5a**. From this viewpoint, it is not surprising that the shear moduli measured with different alkanes are similar, as they are predominantly determined by the intermolecular forces acting between the neighboring molecules at the grain boundaries. Note that, if the chemical nature of the molecules is changed, the intermolecular forces would also change and the moduli would be different.

### 3.3.2 Rotator phases

In this subsection we compare the moduli measured for the different types of R phases and discuss their dependence on the alkane chain-length, absolute temperature and the subcooling from the alkane melting temperature. As seen from the data presented in **Figures 2** and **3**, the R phase moduli depend significantly on the specific temperature at which they are measured. In all cases, the measured moduli increase when the temperature in the sample is decreased. Notably, this increase is usually much smaller compared to the increase accompanying the phase transitions between two distinct R phases. The direct comparison between the dynamic moduli measured for a given R phase in alkanes of different chain lengths is further hampered by the fact that the number of rotator phases appearing in the various alkanes is different.

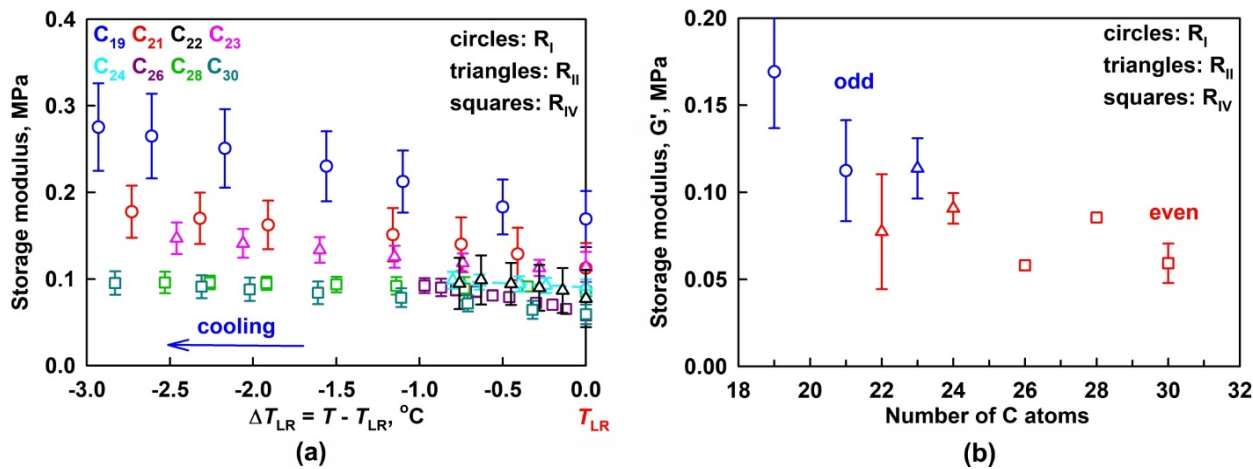

**Figure 6. Storage modulus of the first rotator phase formed from melt upon alkane cooling. (a)** $G'$ as a function of the subcooling $\Delta T_{LR}$ from the temperature of liquid-to-rotator phase transition. $T_{LR}$ was defined from the rheological experiments as the temperature at which the L→R transition has ended (see the dashed lines in **Figures 2** and **3**). The different colors represent the different alkanes as shown on the graph. **(b)** The initial $G'$ values measured for different alkanes just after the end of the L→R transition, i.e. at sobcooling $\Delta T_{LR} = 0$ as defined in (a). The blue symbols show data for the odd-numbered alkanes and the red symbols – for the even-numbered alkanes. The different types of symbol in (a,b) show the different R phases formed: circles – $R_I$, triangles – $R_{II}$ and squares – $R_V$.

Therefore, to compare the moduli of the different R phases, we prepared graphs using a similar approach as applied above for the C phases. In this case, we plotted the moduli measured for the first rotator phase which is observed upon cooling from melt, as a function of its subcooling, $\Delta T_{LR} = (T - T_{LR})$, from the temperature of the L→R transition, $T_{LR}$. Note that the type of this first rotator phase depends on the alkane chain length – it is $R_I$ for $n \leq 21$, $R_{II}$ for $22 \leq n \leq 25$, $R_{IV}$ for $26 \leq n \leq 30$ and $R_{III}$ for $n \geq 31$ [6]. The results from this comparison are presented in **Figure 6a**.

The general trend observed in **Figure 6a** is different from the one observed for the C phases, **Figure 4**. For the rotator phases, we observed that both the storage and loss moduli decrease with the increase of the alkane chain-length. For example, the storage modulus for $C_{19}$ just after the L→R transition was $0.17 \pm 0.03$ MPa at $\Delta T_{LR} = 0°C$, whereas it was ≈ 3 times lower for $C_{30}$, $G' \approx 0.06 \pm 0.01$ MPa. The different R phases exist in different temperature intervals and, therefore, the data points for different alkanes end up at different subcoolings. Nevertheless, the relative difference between the data for the different alkanes remains similar, reflecting the linear increase of the moduli with the temperature decrease. Similar trends were observed for the loss moduli as well.

To refine the comparison of the storage moduli for the different alkanes, , we show in **Figure 6b** the storage moduli of the first R phase observed upon cooling from melt, as measured immediately after its formation at $\Delta T_{LR} = 0°C$. Our data set includes three odd-numbered alkanes with chain-lengths of 19, 21 and 23 C-atoms and five even-numbered alkanes with chain-lengths between 22 and 30 C-atoms. The values for $C_{19}$ were significantly higher than those for the other alkanes, whereas they were practically the same for $C_{21}$ and $C_{23}$ alkanes, $G' \approx 0.11 \pm 0.03$ MPa. For the even-numbered alkanes, the storage modulus for the first rotator phase observed upon cooling at $\Delta T_{LR} = 0°C$ was $G' \approx 0.075 \pm 0.015$ MPa. However, a non-linear trend larger than the experimental error is seen, as the initial storage modulus for $C_{24}$ and $C_{28}$ were very similar, as well as those for $C_{26}$ and $C_{30}$. The interpretation of this trend is currently unknown and further studies, combining the rheological measurements with structural characterization of the studied samples, would be needed to explain it.

The fact that the longer alkanes have smaller dynamic moduli is most probably related to the structural defects which are known to be present within the rotator phases. The most abundant conformers for hydrocarbon chains are *all trans*, *end-gauche*, *kink* and *double-gauche*, see **Supplementary Figure S7** [68-70]. While almost all of the molecules in the crystalline phase are in *all trans* conformation, in the R phase the nonplanar conformers play an important role [69]. The experiments with positron annihilation lifetime spectroscopy (PALS) show that the ortho-positronium lifetime $\tau_3$, characterizing the free-volumes near the *kink* conformers, increases from 2.4 ns for $C_{19}$ to 3 ns for $C_{26}$, whereas it is about 1.1-1.5 ns for the crystalline phases [70]. Respectively, the percentage of "kinked" molecules was assessed to be about 8% in $C_{17}$ and increased to nearly 70% for $C_{29}$ [68,70]. Therefore, the observed "softening" of the rotator phases is most probably caused by increased fraction of nonplanar molecules which disturb the molecular ordering. Our results for the crystalline phase moduli are also in a good agreement with the studies showing that the positronium lifetime $\tau_3$ does not depend on the alkane chain length in crystalline phase [70].

In conclusion, the moduli of the rotator phases are about one order of magnitude smaller than those of the crystalline phases, reflecting the "softness" of the plastic rotator phases. The moduli of the crystalline phases do not depend on the alkane chain-length (in the range of tested alkanes). Therefore, they are most probably related to the ease of sliding between the separate crystalline domains in the polycrystalline structure. The main fraction of the molecules in the C phases is in the *all trans* conformation. In contrast, the rotator phase moduli depend on the alkane chain-length and lower values are measured with longer alkanes. The latter trend is explained with the increased fraction of molecules present in nonplanar conformations, which is known to increase significantly with the alkane chain-length.

**3.4 Applicability of the rheological protocol to samples of more complex composition.**

Bulk rotator phases have been observed to form in asymmetric alkanes, 1-alkenes, long chain alcohols, natural waxes and various mixtures of such substances [9]. Our limited experiments with $C_{20:1}$ alkene, using the same measurement protocol, showed that the L→R→C transitions can be also detected with rheological measurements, see **Supplementary Figure S8**. Because the focus of the current study are the various alkanes, in this section we present illustrative results about the viscoelastic properties of several alkane mixtures and compare them to those obtained with pure alkanes.

Many studies have shown that mixing two alkanes significantly expands the temperature interval of existence of the rotator phases. The $R_I$ phase observed in $C_{17}$ and $C_{19}$ pure alkanes exists within 6-8°C temperature interval, whereas their 1:1 mixture forms rotator phase which is observed in an interval of 24°C [27]. Furthermore, alkane mixing leads to increased rotator phase stability for the even-numbered alkanes which form only short living transient phases when studied in pure form [11]. For example, in 1:1 $C_{16}:C_{18}$ mixture, the $R_I$ phase forms at *ca.* 18°C and exists down to -5°C. It was found, however, that the structural and thermodynamic properties of such mixed rotator phase do not remain the same in the entire temperature range of existence, but change and become closer to those of the crystalline phases upon temperature decrease [25].

To check how the described behavior is reflected in the rheological properties of alkane mixtures, we performed series of measurements with several alkane blends. The phase diagram of the phases formed in $C_{17}$-$C_{19}$ mixtures of different ratios is presented in **Supplementary Figure S9a** and illustrative example for the rheological results obtained with such $C_{17}$-$C_{19}$ mixtures is presented in **Figure 7a**. For all tested binary mixtures, the L→R phase transition occurs as expected, with the moduli increasing up to $10^4$-$10^5$ Pa. In the R phase, the moduli increased upon further cooling. However the R→C transition in $C_{17}$-$C_{19}$ mixtures occurred with a change of the curve slope, instead of the stepwise increase as observed with the pure alkanes, see **Figure 7a**. This is most probably related to the relatively small (incremental) structural changes in the sample upon R→C transition, due to the significantly expanded area of existence of the R phase (in this sample it is observed upon cooling from 27.7°C down to 6.9°C).

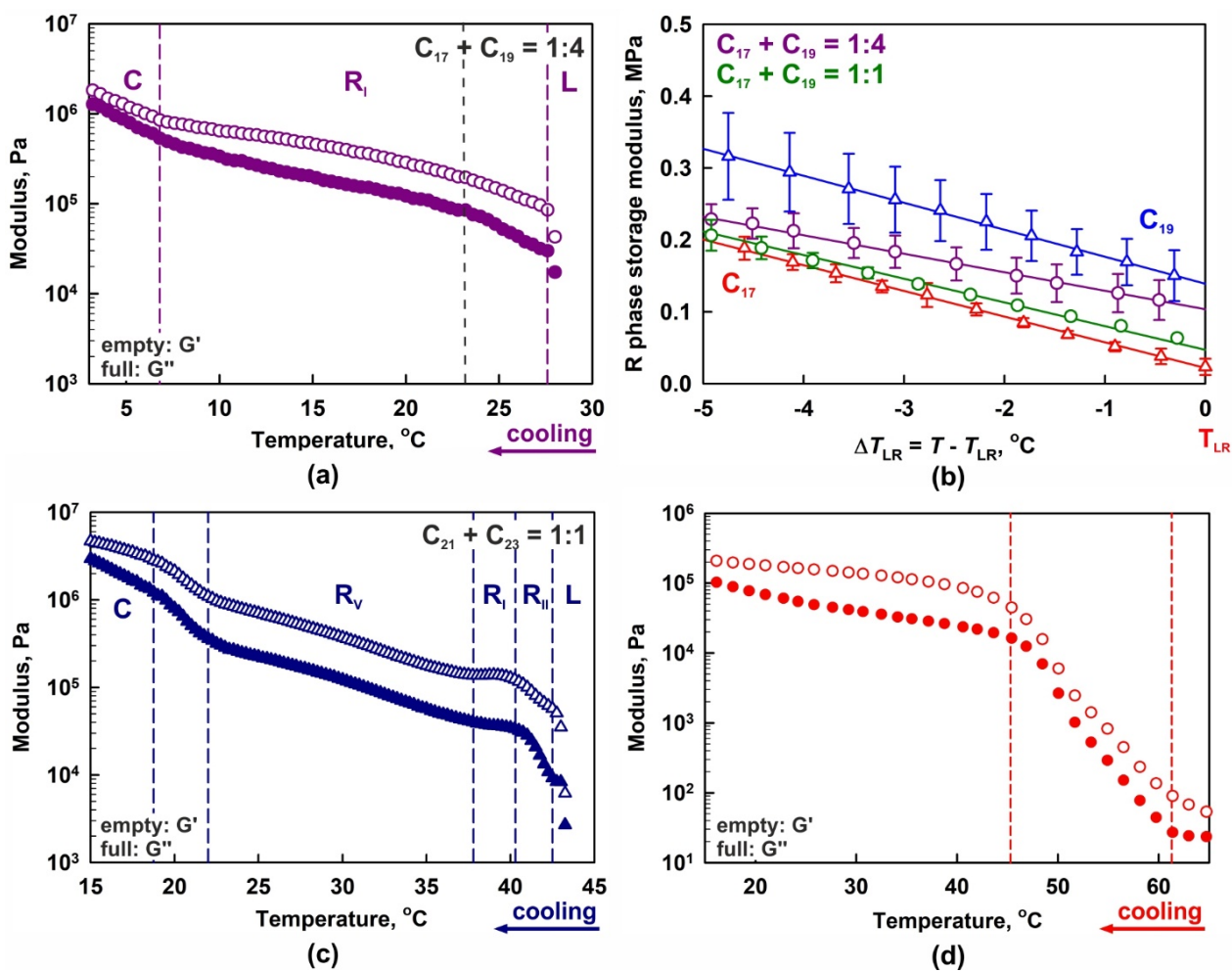

**Figure 7. Rheological properties of alkane mixtures.** Storage and loss moduli measured in oscillatory shear experiments for: **(a)** $C_{17} + C_{19}$ = 1:4 mixture; **(c)** $C_{21} + C_{23}$ = 1:1 mixture and **(d)** commercial Vaseline sample. **(b)** Summarized dependence of the $G'$ modulus for R phases observed in nonadecane, heptadecane and two $C_{17}$-$C_{19}$ mixtures, as a function of the subcooling from the liquid-to-rotator phase transition temperature, $T_{LR}$. All experiments are performed upon cooling with 0.5°C/min rate from melted samples. Empty symbols show the storage modulus ($G'$) and the full symbols – the loss modulus ($G''$).

Another interesting observation for the $C_{17}$:$C_{19}$ = 1:4 sample was that the slope of the curves moduli vs. temperature changed at about 23.6°C, **Figure 7a**. As explained in the previous sections, such slope changes are observed when phase transition between phases with similar characteristics occurs. The temperature of 23.6°C is very close to the melting temperature of the bulk heptadecane ($T_m \approx 22$°C), therefore it is possible that the slope change is due to further solidification of heptadecane-enriched domains. However, no additional phases beside $R_I$ have been reported currently in the literature for this mixture. Therefore, additional structural SAXS/WAXS measurements are needed to resolve the molecular arrangement and its changes in this temperature interval.

Measuring the $R_I$ phase moduli for three $C_{17}$-$C_{19}$ mixtures of different ratios, along with the data that we had for the pure compounds, allowed us to compare the properties of the $R_I$ phase as a function of the nonadecane fraction in the mixture. The results for the storage moduli of the different mixtures are presented in **Figure 7b** as a function of the subcooling of the R phase from the melt, $\Delta T_{LR} = T - T_{LR}$. The results for the $C_{17}$:$C_{19}$ = 1:9 mixture are not shown, as they were practically the same (in the frame of our experimental accuracy) to those for $C_{17}$:$C_{19}$ = 1:4 mixture. This comparison shows that the moduli of the mixtures have intermediate values between the moduli of the respective pure compounds. The increase of the $C_{17}$ fraction in the mixture decreases the measured moduli because the $G'$ values for pure $C_{17}$ are significantly lower than those for $C_{19}$.

The rheological response of $C_{21}$ + $C_{23}$ mixture in 1:1 molar ratio was also measured as an example for mixture undergoing three different R phases upon cooling. The phase diagram known from the literature is presented in **Supplementary Figure S9b** [27] and the obtained results from the rheological measurements are presented in **Figure 7c**. Once again, the moduli dependence on the temperature allows us to detect all phase transitions occurring in the sample. In contrast to $C_{17}$ + $C_{19}$ mixture, in this case the R→C phase transition occurs with finite stepwise increase of the measured moduli.

As representative of highly mixed alkane system, we studied the rheological properties of commercially sold Vaseline, which is a mixture of alkanes with chain lengths between *ca.* 14 and 50 C-atoms. The cooling DSC thermogram of the studied Vaseline sample shows that the freezing process starts at ≈ 66°C and has not ended completely down to 0°C, see **Supplementary Figure S10**. Therefore, the R→C phase transition was not observed in the rheological measurements, due to the very low temperature of its occurrence. The L→R phase transition, however, is easily seen in **Figure 7d**. In this case, the transition occurred in a very wide temperature interval, between *ca.* 60°C and 45°C. The 45°C temperature coincides with the maximum of the exothermic peak observed in the DSC thermogram. Once again, the $G'$ modulus was higher than the $G''$ modulus and, when the R phase was formed, the moduli were in the range of $10^4$ to $10^5$ Pa and slightly increased upon further cooling.

These illustrative results prove that the rheological measurements can be successfully applied to many different systems forming intermediate rotator phases. Our results for the binary $C_{17}$-$C_{19}$ mixtures suggest that the rheological properties of the mixtures will be intermediate between those for the respective pure compounds. However, further dedicated experiments need to be performed to confirm or reject this hypothesis with a larger number of various mixtures.

## 4. Conclusions

The current study explores systematically the shear rheological properties of the thermodynamically stable crystalline rotator phases and of the fully crystalline phases of normal alkanes with chain-lengths varied between 17 and 30 carbon atoms. We show that the storage ($G'$) and loss ($G''$) moduli measured in oscillatory shear experiments performed at 0.05 % strain and 1 Hz frequency can be successfully used to detect all phase transitions occurring in these samples upon both cooling and heating.

The obtained results show that the moduli measured for crystalline phases of alkanes with different chain lengths do not differ from one another, despite the fact that the molecules arrange in different crystalline lattices. In contrast, the moduli for the first rotator phase observed upon alkane cooling decreased their values when the alkane chain length increases.

We hypothesize that these trends can be explained assuming that the $G'$ and $G''$ values are determined mainly by the friction between the distinct crystalline domains in the polycrystalline structure, while in the rotator phases we have larger fraction of kinks (defects) in the alkanes with longer chains which softens the respective plastic structures. Indeed, previous studies report no significant difference in the fraction of defects in the crystalline phases of the different alkanes, whereas a significant increase in the number of defects was reported for the rotator phases with the increase of alkane chain length [68-70].

The measured moduli depend on the absolute temperature and on the subcooling of the alkane from its melt – higher moduli were measured at lower temperatures reflecting the mechanical strengthening of the structure, most probably caused by the decreased thermal energy and the tighter molecular packing in the system. The storage moduli of the R phases were found to vary in the range between 0.1 and 1 MPa, while the loss moduli were about 3-4 times lower. For the crystalline phases, $G'$ was in the range between 0.5 and 3 MPa, depending on the subcooling, while the loss moduli were in the range between 0.2 and 0.5 MPa.

We show also that the rheological measurements can be used to detect and characterize the phase transitions in bulk alkenes and alkane mixtures. Therefore, the proposed experimental procedure and its variations (e.g. measurements at different frequencies and amplitudes) are expected to be applicable to wide range of materials forming intermediate plastic phases between their crystalline and liquid phases.

**Acknowledgements:**

The study was funded by the Bulgarian Ministry of Education and Science, project KP-06-COST-10. The study falls under the umbrella of European Network COST CA 17120. The authors thank Ms. Hristina Bletsova and Ms. Aleksandra Todorova (Sofia University) for performing some of the rheological measurements.

**Author contributions:**

D.C. – Conceptualization, Validation, Visualization, Writing – original draft, review & editing

K.T. – Investigation, Validation

S.T. – Conceptualization, Methodology, Supervision, Writing – review & editing

N.D. – Conceptualization, Methodology, Supervision, Funding acquisition, Writing – review & editing

# Supplementary materials

# Rheological properties
# of rotator and crystalline phases of alkanes

**Diana Cholakova, Krastina Tsvetkova**

**Slavka Tcholakova, Nikolai Denkov***

*Department of Chemical and Pharmaceutical Engineering*

*Faculty of Chemistry and Pharmacy, Sofia University,*

*1 James Bourchier Avenue, 1164 Sofia, Bulgaria*

*Corresponding author:

Prof. Nikolai Denkov

Department of Chemical and Pharmaceutical Engineering

Sofia University

1 James Bourchier Ave.,

Sofia 1164

Bulgaria

E-mail: nd@lcpe.uni-sofia.bg

Tel: +359 2 8161639

Fax: +359 2 9625643

**Supplementary Table S1.** Alkanes used in the study. The purity and melting temperatures are given as provided by the producers.

|  | Notation in text, $C_n$ | Producer | Purity, % | $T_m$, °C |
|---|---|---|---|---|
| **Heptadecane** | $C_{17}$ | Sigma-Aldrich | 99 | 20-22 |
| **Nonadecane** | $C_{19}$ | TCI Chemicals | ≥ 99 | 32 |
| **Eicosane** | $C_{20}$ | Sigma-Aldrich | 99 | 35-37 |
| **Heneicosane** | $C_{21}$ | TCI Chemicals | ≥ 99 | 41 |
| **Docosane** | $C_{22}$ | TCI Chemicals | ≥ 98 | 46 |
| **Tricosane** | $C_{23}$ | Sigma-Aldrich | 99 | 46-47 |
| **Tetracosane** | $C_{24}$ | TCI Chemicals | ≥ 99 | 51 |
| **Hexacosane** | $C_{26}$ | Sigma-Aldrich | 99 | 55-58 |
| **Octacosane** | $C_{28}$ | Sigma-Aldrich | 99 | 57-64 |
| **Triacontane** | $C_{30}$ | Sigma-Aldrich | 98 | 64-67 |

**Supplementary Table S2.** Comparison between the phase transition temperatures for $C_{21}$ alkane as measured in our DSC and rheological experiments upon cooling (in blue) and upon heating (in red). Note that when the sample is heated the $C \to R_I \to L$ phase transitions take place, whereas upon cooling the transitions are reversed, $L \to R_I \to C$.

|  | Temperatures measured *upon cooling*, °C ||||| Temperatures measured *upon heating*, °C |||||
|---|---|---|---|---|---|---|---|---|---|---|
| Method | DSC ||| Rheology || DSC ||| Rheology ||
| Phase transition | $T_{start}$ | $T_{max}$ | $T_{end}$ | $T_{start}$ | $T_{end}$ | $T_{start}$ | $T_{max}$ | $T_{end}$ | $T_{start}$ | $T_{end}$ |
| $L \rightleftarrows R_I$ | 40.0 | 39.7 | 39.2 | 41.3 | 40.3 | 39.7 | 40.7 | 42.3 | 41.0 | 42.1 |
| $R_I \rightleftarrows C$ | 30.8 | 30.7 | 29.8 | 32.1 | 30.6 | 31.7 | 32.6 | 33.5 | 32.3 | 33.8 |

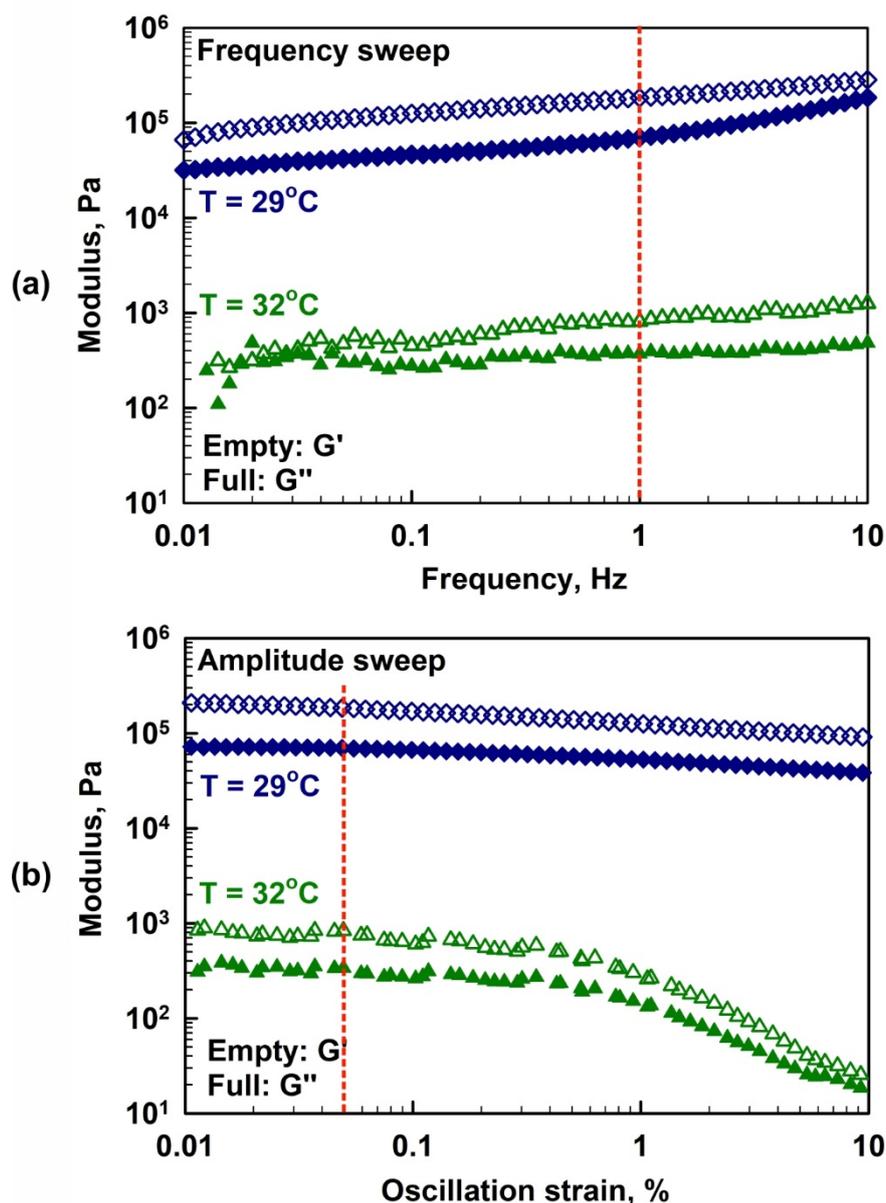

**Supplementary Figure S1. Results from dynamic oscillation rheological tests performed with nonadecane ($C_{19}$) samples at different constant temperatures.** (a) Frequency sweep experiments performed at 0.05 % strain. (b) Amplitude sweep experiments performed at 1 Hz frequency. The sample has been placed at 35°C and cooled down to the desired temperature at 0.5°C/min rate. Note that 32°C is just at the onset when the liquid to rotator phase transition begins, therefore, the obtained results differ significantly from the others. Empty symbols show the storage modulus ($G'$) and the full symbols – the loss modulus ($G''$). The red dashed lines show the frequency and strain values, chosen for the main fraction of the oscillatory experiments performed upon cooling or heating.

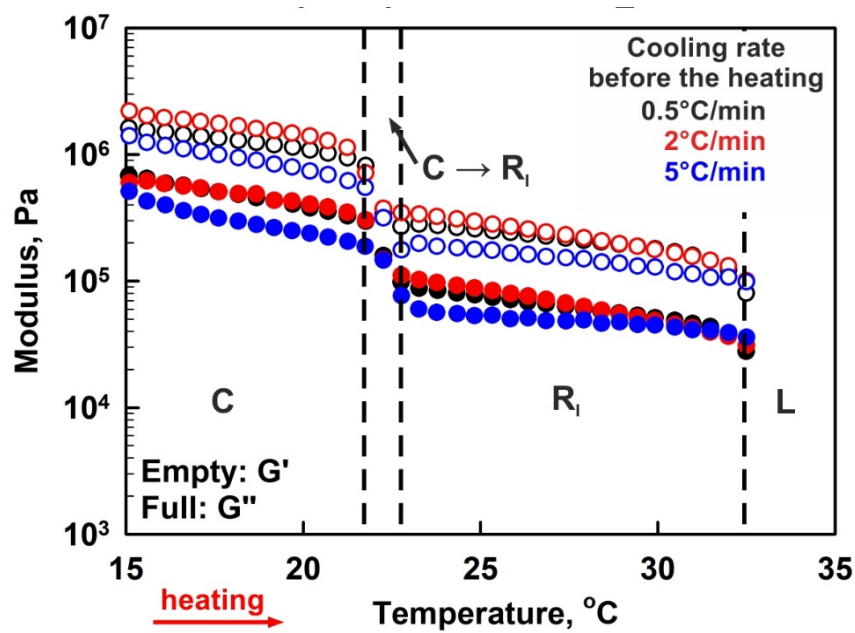

**Supplementary Figure S2.** **Dynamic shear moduli measured with bulk $C_{19}$ samples upon heating**. The samples have been cooled from melt at different cooling rates as indicated by the colours on the graph: black – 0.5°C/min; red – 2°C/min and blue – 5°C/min. All samples are heated at 0.5°C/min. Empty symbols show the storage modulus ($G'$) and the full symbols – the loss modulus ($G''$). The results obtained at 0.5°C/min and 2°C/min cooling rates are within the frame of our experimental accuracy. At 5°C/min rate we obtained slightly lower moduli showing a dependence of the mechanical properties of the crystals from the cooling. Similar trends have been previously demonstrated for different waxes [1,2].

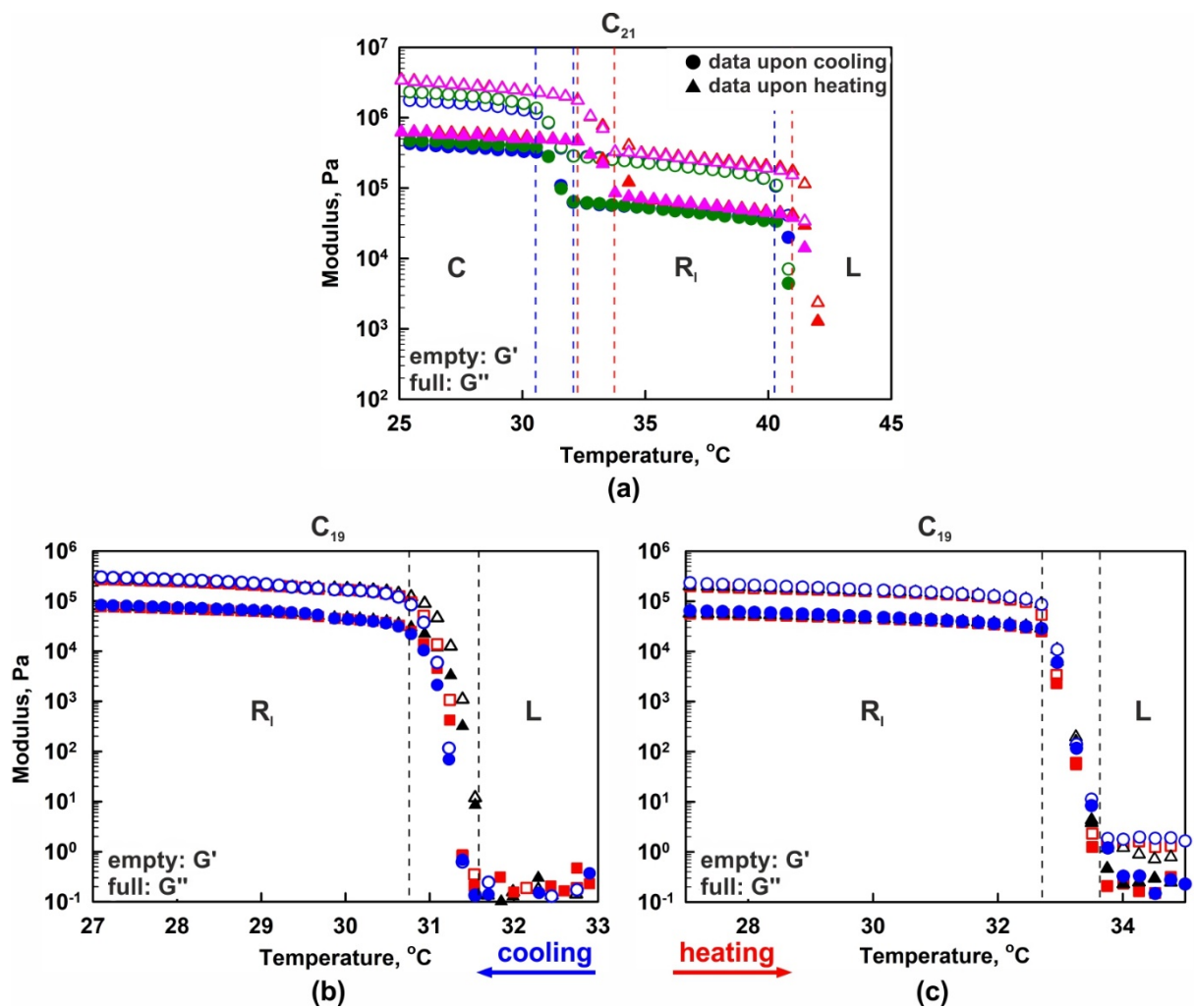

**Supplementary Figure S3.** **Data from rheological measurements with: (a) Heneicosane, $C_{21}$** and **(b-c) Nonadecane, $C_{19}$**. (a) Comparison between cooling (circles) and heating (triangles) curves. The measured moduli are in a good agreement with one another, however, the phase transition temperatures are slightly shifted as discussed in the main manuscript. (b) Data obtained from melt upon cooling to 27°C. R phase is present at the end of the experiment. (c) Data obtained upon heating from 27°C until melting. All experiments are performed at 0.5°C/min cooling and heating rates. Empty symbols show the storage modulus ($G'$) and the full symbols – the loss modulus ($G''$).

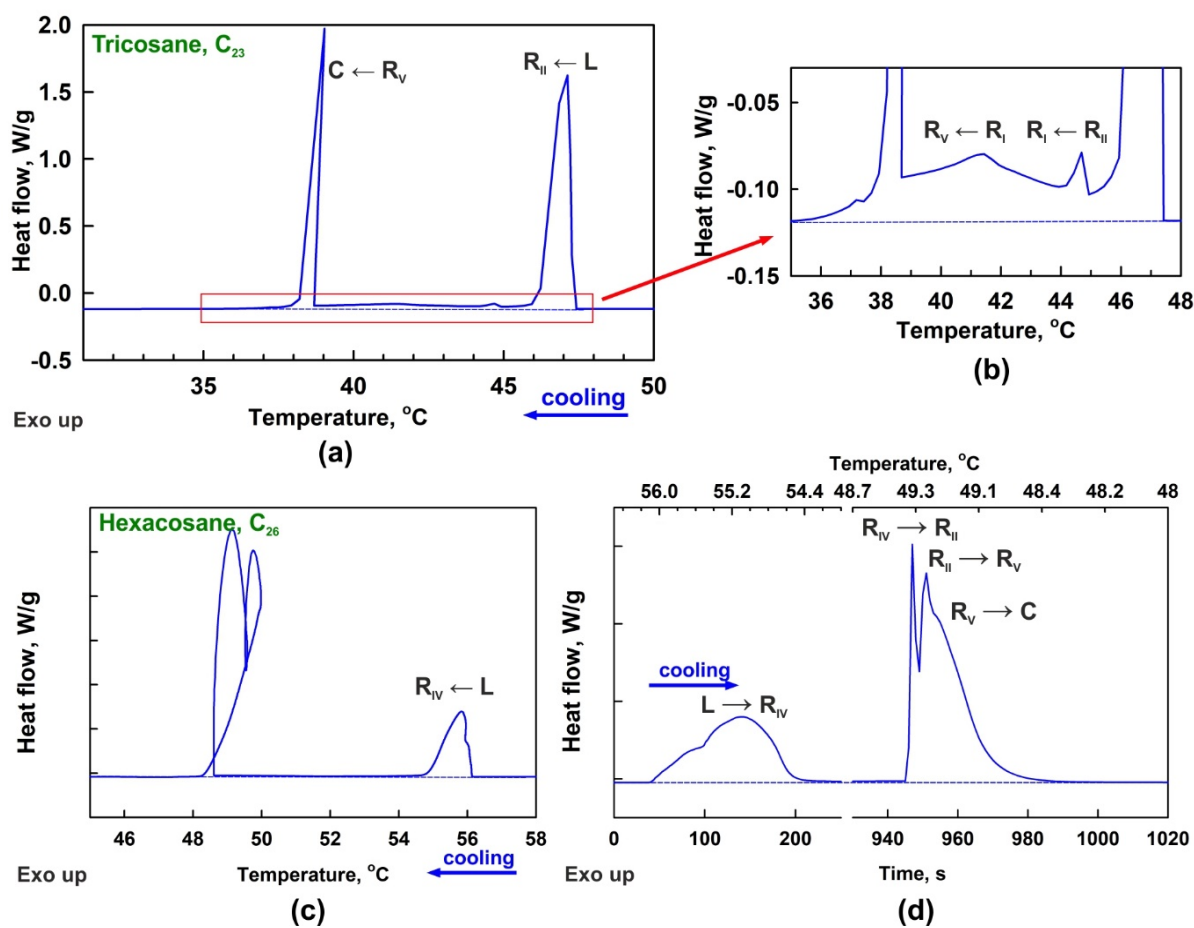

**Supplementary Figure S4.** DSC thermograms upon 0.5°C/min cooling of $C_{23}$ (a,b) and $C_{26}$ (c,d). **(a-b)** The main peaks representing the L→$R_{II}$ and $R_V$→C phase transitions are visible at the large heat flow scale as shown in (a), whereas the phase transitions between the distinct R phases are shown in the zoom-in view in (b). **(c)** The main peaks observed at the Q(T) graph for $C_{26}$ are asymmetrical because they show several consecutive phase transitions. The loop observed in the low temperature peak (47°C) is due to the very high heat released in the sample, leading to its temporary temperature increase. When the same data is plotted as a function of the time, instead of temperature, as shown in **(d)**, the data shows that the peak with maximum at ≈ 54°C is caused by two consecutive phase transitions, while the low-temperature peak observed at ca. 47°C is caused by three consecutive phase transitions, as indicated with the labels in (d).

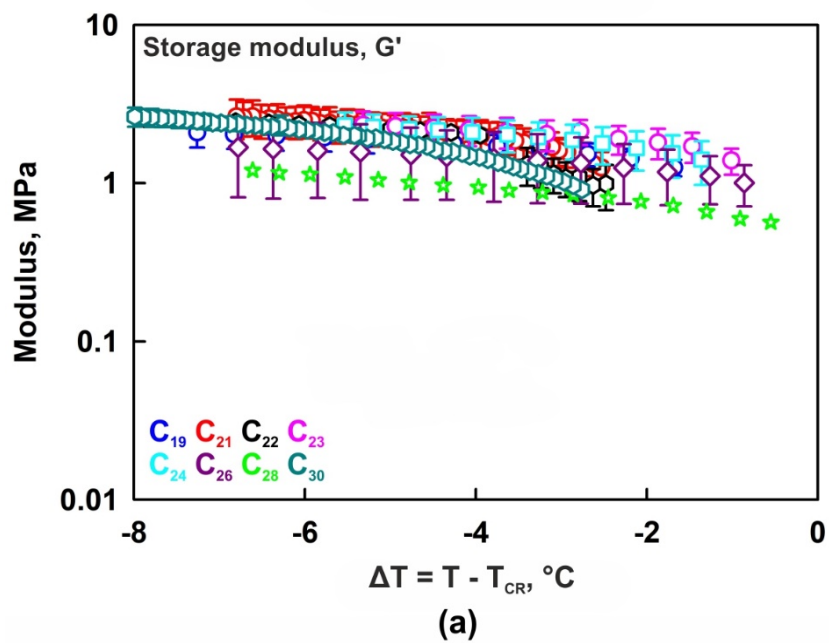

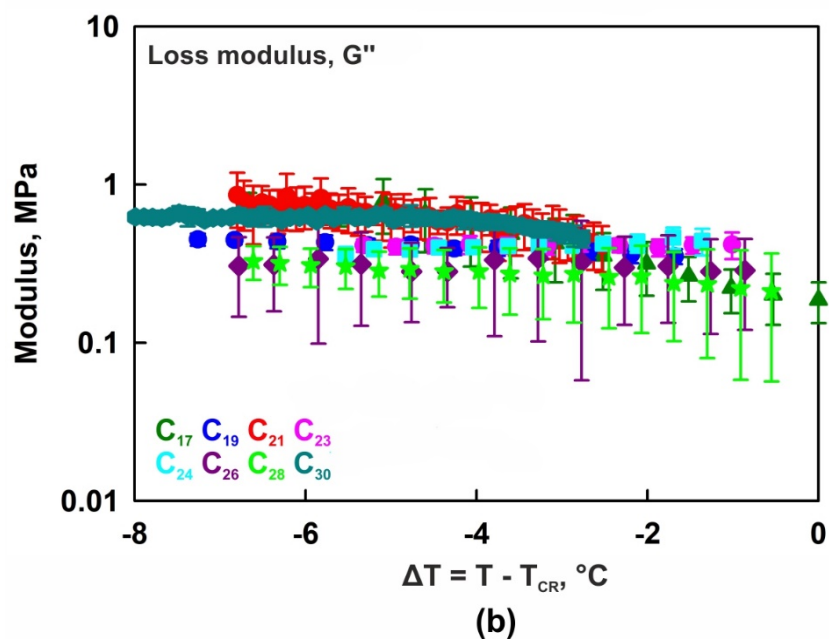

**Supplementary Figure S5. Storage (a) and loss (b) moduli of the crystalline phases for alkanes with different chain lengths.** The data is obtained upon cooling. The subcooling $\Delta T_{CR}$ on *x*-axis is calculated with respect to the temperature of the crystal to rotator phase transition observed upon heating to be directly comparable to the *x*-axis shown on **Figure 4** in the main text.

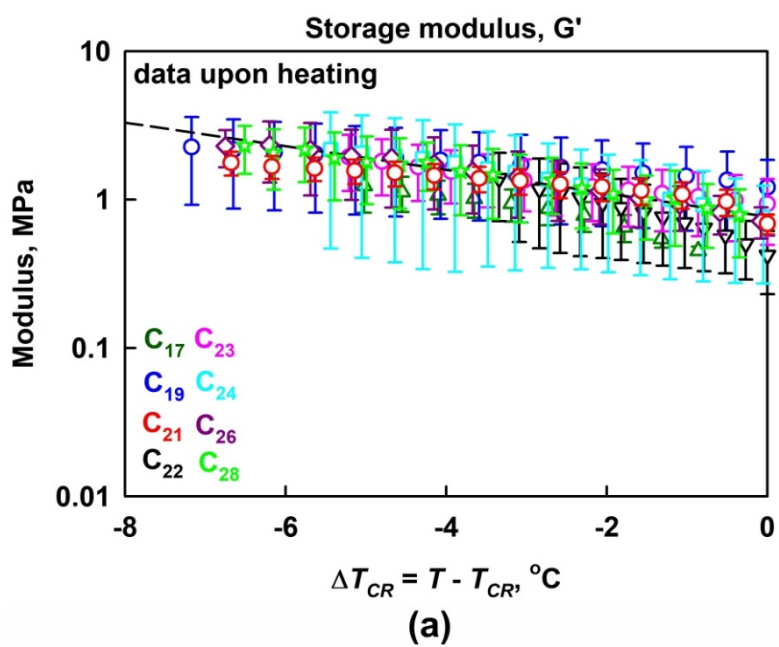

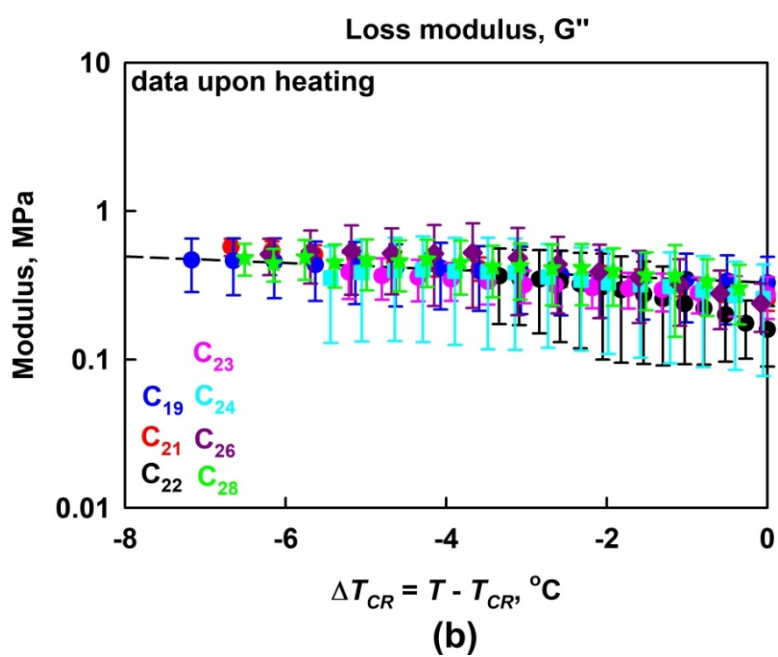

**Supplementary Figure S6.** **Storage (a) and loss (b) moduli of crystalline alkane phases.** The moduli are plotted as a function of the subcooling $\Delta T_{CR}$ with respect to the temperature at which the C→R phase transition occurs upon heating. Data for all alkanes with $n \geq 19$ lay on master lines, showing that the shear moduli do not depend significantly on the type of the crystal lattice formed.

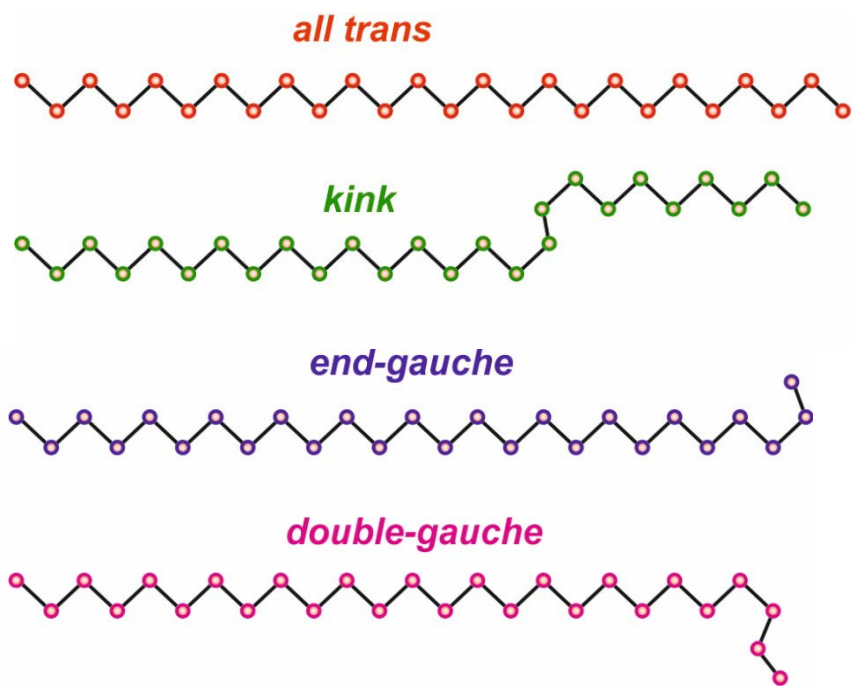

**Supplementary Figure S7.** Schematic presentation of the most abundant **hydrocarbon chain conformers**. The hydrogen atoms are omitted for simplicity. Adapted from Ref [3].

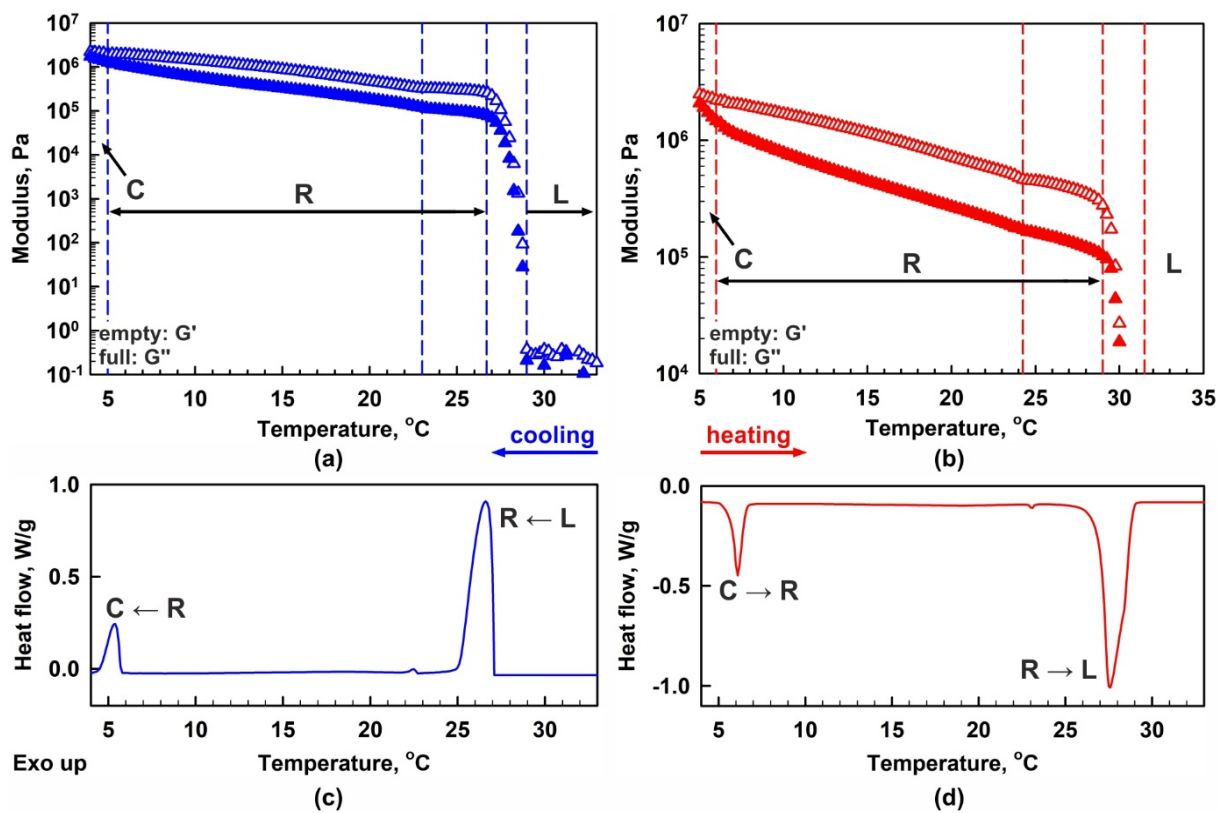

**Supplementary Figure S8. Rheological (a,b) and DSC (c,d) measurements of 1-eicosene.** Data obtained upon cooling is shown in (a,c) and the curves measured upon heating in (b,d). Note that the small peak observed in the thermograms at ≈ 22.5-23°C is reflected in the rheological measurements with a slope change. The R→C and C→R phase transitions also occur with a slope change in the rheological measurements. All experiments are performed at 0.5°C/min rate. Empty symbols show the storage modulus ($G'$) and the full symbols – the loss modulus ($G''$).

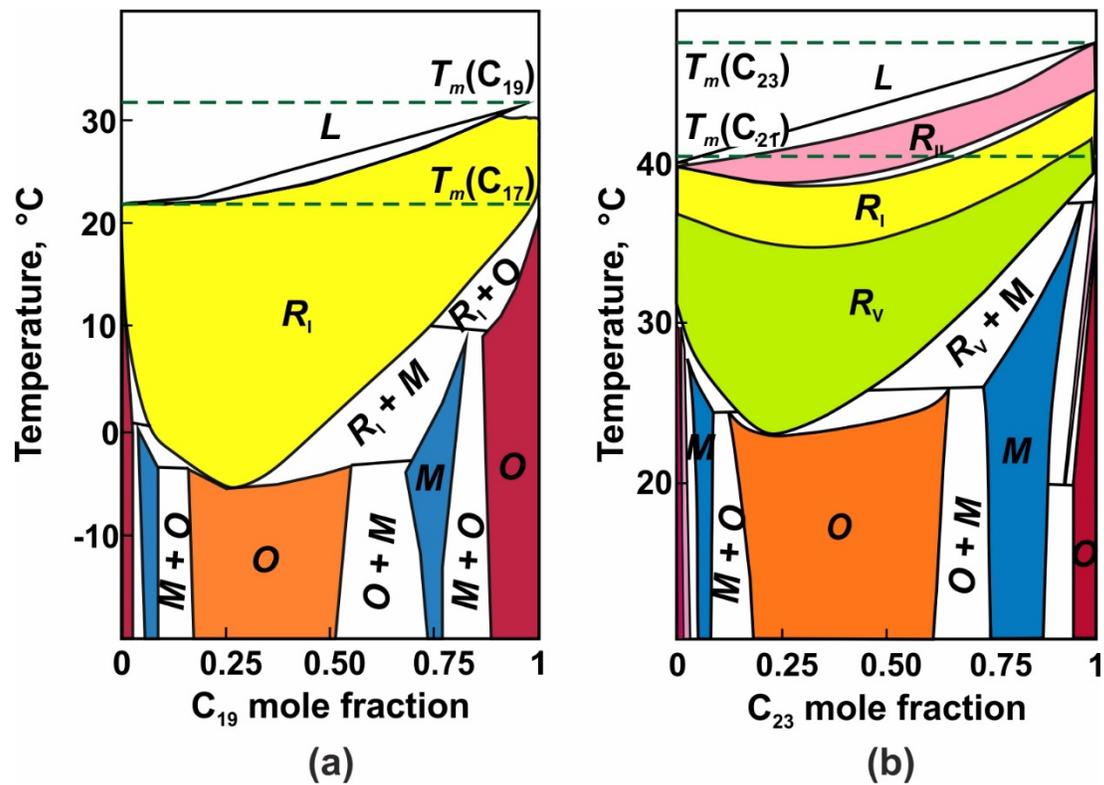

**Supplementary Figure S9.** Binary phase diagrams for alkane mixtures: (a) $C_{17} - C_{19}$ mixture; (b) $C_{21} - C_{23}$. Adapted from Refs [4,5].

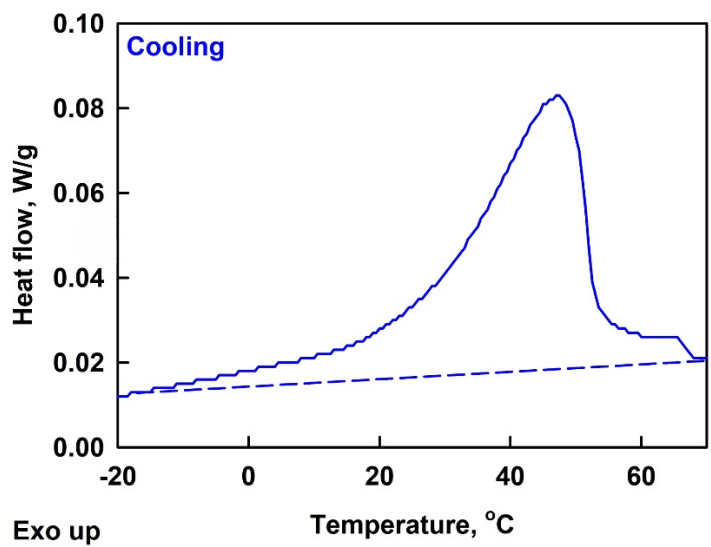

**Supplementary Figure S10. DSC thermogram of Vaseline.** The thermogram is obtained upon cooling at 2°C/min from melted sample. Note the wide temperature interval in which the freezing occurs – it starts at 67.5°C and ends at ca. -20°C. The observed peak represents a combination of both liquid to rotator and rotator to crystal phase transitions.